Version 7

# Systemic lupus erythematosus in African-American Women: Cognitive physiological modules, autoimmune disease, and structured psychosocial stress

Rodrick Wallace, Ph.D.
The New York State Psychiatric Institute*

November 20, 2003

## Abstract

Examining elevated rates of systemic lupus erythematosus in African-American women from perspectives of immune cognition suggests the disease constitutes an internalized physiological image of external patterns of structured psychosocial stress, a 'pathogenic social hierarchy' involving the synergism of racism and gender discrimination, in the context of policy-driven social disintegration which has particularly affected ethnic minorities in the USA. The disorder represents the punctuated resetting of 'normal' immune self-image to a self-attacking 'excited' state, a process formally analogous to models of punctuated equilibrium in evolutionary theory.

Thus disease onset takes place in the context of a particular immunological 'cognitive module' similar to what has been proposed by evolutionary psychologists for the human mind. Disease progression involves interaction of the hypothalamic-pituitary-adrenal axis, which we also treat as a cognitive physiological submodule, with both immune cognition and an embedding pathogenic social hierarchy, a structured psychosocial stress which literally writes an image of itself on the course of the disorder. Both onset and progression may be stratified by a relation to cyclic physiological responses which are long in comparison with heartbeat period: circadian, hormonal, and annual light/temperature cycles.

The high rate of lupus in African-American women suggests existence of a larger dynamic which entrains powerful as well as subordinate population subgroups, implying that the wide ranging programs of social and economic reform required to cause declines in disease among African-American women will bring significant benefit to all.

## Introduction

Systemic lupus erythematosus (SLE) is a multisystem autoimmune disorder which most frequently affects young women. Arthritis, skin rash, osteoporosis, cataracts, accelerated atherosclerotic vascular disease (ASVD), central nervous system dysfunction, and renal disease are the most common manifestations, whose severity may markedly fluctuate over time. The damage of the disease is of 'Type III', i.e. mediated by immune complexes which can range from just a few molecules to relatively huge structures involving whole cells coated or cross-linked by antibody, accounting for the great variety of pathology seen in this form of illness (Paul, 1999; Liang et al., 2002). The disease is characterized by polyclonal B-cell activation, elevated production of pathogenic autoantibodies, impaired immune complex clearance and inflammatory responses in multiple organs. Like asthma, the pathological cascade is marked by an imbalance between depressed Th1 cell cytokines, which promote cell-mediated immunity, and enhanced Th2 cell cytokines, which support humoral immunity. There increasingly strong evidence that the cytokine Interleukin-6 (IL-6) is central to this process. IL-6 is a B-cell differentiation factor that induces the final maturation of IL-4-preactivated B cells into immunoglobulin (Ig)-secreting plasma cells (e.g. Schotte et al., 2001; Linker-Israeli et al., 1991; Cross and Benton, 1999).

Kiecolt-Glaser et al. (2002) discuss how chronic inflammation involving IL-6 has been linked with a spectrum of conditions associated with aging, including cardiovascular disease, osteoporosis, arthritis, type II diabetes, certain cancers, and other conditions. In particular the association between cardiovascular disease and inflammation, as mediated by IL-6, is related to its central role in promoting the production of C-reactive protein (CRP), an ancient and highly conserved protein secreted by the liver in response to trauma, inflammation, and infection.

CRP is a pattern recognition molecule of the innate immune system keyed to surveillance for altered self and certain pathologies, providing early defense and activation of the humoral, adaptive, immune system. It is increasingly seen as a linkage between the two forms of immune response (Du Clos, 2000; Volanakis, 2001).

As Cross and Benton (1999) note, although IL-6 (and IL-10) have been most intensely studied for involvement in the pathogenesis of SLE, the cascade nature of cytokines means that all components of the cytokine network must, ultimately, be considered. We shall attempt to model this in a very general way below.

Within the US, SLE disproportionately affects African-

*Address correspondence to R. Wallace, PISCS Inc., 549 W 123 St., Suite 16F, New York, NY, 10027. Telephone (212) 865-4766, email rdwall@ix.netcom.com. Affiliation for identification only.

American women, and accelerated ASVD occurs in subjects who are predominately premenopausal women at an age when ASVD is rare or unusual (Liang et al., 2002; Bongu et al., 2002). Between 1979 and 1998, SLE death rated have increased approximately 70 percent among African-American women aged 45-64 years (MMWR, 2002).

The basic disparity in disease occurrence is considerable, approximately four times higher in African-Americans than Caucasians (Bongu et al., 2002). Among Caucasian women, total SLE mortality has remained stable since the late 1970's at about 4.6 deaths per million with a decline in rates in younger and a rise in older women. Among African-American women, total SLE mortality rose 30 percent to a mean annual rate of 18.7 per million, with a constant rate in younger and a rising rate in older women. The rising disparity involves both increasing prevalence and worse disease in younger African-American women (Bongu et al., 2002).

Parks et al. (2002) find that the increased risk of SLE in African-Americans cannot be explained by hormonal or reproductive risk factors (i.e. breastfeeding, preeclamsia), occupational exposures (i.e. silica, mercury), medication allergy, herpes zoster, or similar factors. They suggest, rather, a central role for such "personal and social stressors" as racism and poverty in creating the disparity, a fundamental insight which we will explore at more length below.

Here we will adopt the perspective of the theory of immune cognition (e.g. I.R. Cohen, 2000; Atlan and Cohen, 1998; Hershberg and Efroni, 2001; Tauber, 1998; Grossman, 1989,1992, 1993a, b, 2000). We particularly embrace Cohen's version of the 'immunological homunculus' – an idea first presented by Whittingham and Mackey (1971) – as a cognitive submodule representing the immune system's self-image of the body. Although Cohen takes an idiotype/anti-idiotype network as the basis of the homunculus, other versions are possible (e.g. the thymus theory originally proposed by Mackey; Rose, 2002), and would probably work as well.

We shall use a series of mathematical models to delineate the basic biology of what can be characterized as 'pathogenic social hierarchy' in the etiology of the disease. This is far from a trivial enterprise: we must first produce a general treatment of autoimmune disease, which we then apply specifically to SLE. Following Pielou (1977), however, we must be careful to understand the principal utility of mathematical models is in raising questions for empirical study, rather than answering them, a matter to which we will return.

We assume some familiarity with earlier work in this direction, using an information theory approach (e.g. Cohen, 2000; Atlan and Cohen, 1998; Wallace and Wallace, 2002; Wallace, 2002a).

We will express deviations from a 'zero-order reference state' of the immunological homunculus in terms of a relatively few 'nonorthogonal eigenmodes' representing complex systemic responses to applied perturbation – infection, tumorigenesis, tissue damage, and the like. These eigenmodes – autoimmune address of self-antigens – are a combination of innate and learned responses to such perturbation. The nonorthogonality implies the possibility of plieotropic excitation of several characteristic autoimmune pathologies by a single perturbation.

The essence of the argument is recognition that the immunological homunculus, the immune system's image of the self, is not a simple physical structure whose zero-order reference mode is a minimum energy state to which the system will automatically return, like a collection of weights on springs left to itself: all states of the immune system are, relatively speaking, rather active high energy states. We infer, then, the necessity of a *cognitive* decision by the immune system as to which of the possible nonorthogonal eigenmodes of the immunological homunculus is to be taken as the zero-order-reference mode to which the system is reset, i.e. the 'normal' pattern of self-recognizing maintenance activities of the immune system. This line of reasoning seems analogous to Nunney's (1999) argument regarding the necessity of an elaborate tumor control strategy for large animals, since the probability of tumorigenesis grows synergistically as the 0.4 power of cell count times animal lifetime, itself dependent on animal size. Some similar power law calculation can probably be done comparing the number of possible 'eigenmodes' of the immune homunculus vs. the 'murunculus' for mouse models of autoimmune diseases.

We are, then, proposing that the immune system has evolved a number of interlocking 'cognitive modules' in a manner analogous to that proposed by evolutionary psychologists for the human mind (e.g. Barkow et al. , 1992). Cohen's immunological homunculus self-image of the body is clearly one such module. We will infer the necessity of others.

Cognitive processes have dual information sources which, through Rate Distortion Theorem (RDT) or Joint Asymptotic Equipartition Theorem (JAEPT) arguments, can become linked across levels of organization with external structured 'signals' of one kind or another in a punctuated or 'phase transition' manner. Thus an appropriate signal – an infection, chemical exposure, or pattern of psychosocial stress – can, if strong enough, suddenly reset the zero-order of the immunological homunculus to a mode different from the learned zero-order maintenance mode, i.e. an actively self-attacking mode, in a manner recognizably analogous to the Eldredge/Gould model of evolutionary punctuation (e.g. Eldredge, 1985; Gould, 2002; Wallace, 2002b). The (relatively) limited number of possible high probability activated states – nonorthogonal eigenmodes – then, accounts for the limited number possible autoimmune diseases. Different excited eigenmodes will generally be triggered by different patterns of external signals through the cognitive reset-to-zero process.

We shall be particularly interested in the possible role of pathogenic social hierarchy as such a signal in the onset of SLE.

Onset of lupus is, in this model, followed by particular developmental pathways of disease expression modulated by the hypothalamic-pituitary-adrenal (HPA) axis, the 'flight-or-fight' mechanism. The HPA axis, through production of adrenal glucocorticoids, can upregulate anti-inflammatory and downregulate pro-inflammatory cytokines, encouraging Th2 at the expense of Th1, immune phenotype. HPA axis activity can, then, serve to turn off inflammatory activation. We shall model the HPA axis as a cognitive submodule which can interact synergistically with both the reset-to-zero module and embedding psychosocial stress to produce disease.

To paraphrase Sternberg (2001), both excess or inadequate stress hormone responses by the HPA are associated with disease. Excessive suppression of Th1 phenotype will enhance susceptibility to infection, while inadequate suppression will enhance susceptibility to inflammatory autoimmune and allergic disease. Thus chronic HPA axis overactivation, as occurs during stress, can affect susceptibility to or severity of infectious disease through the immunosuppressive effects of the glucocorticoids. In contrast, blunted HPA axis responses, for example the consequence of post traumatic stress disorder (PTSD), may enhance susceptibility to autoimmune disease.

We shall model the HPA axis as a cognitive system involving a form of signal transduction driven by the 'magnitude' of average applied stress, much like a stochastic resonance. Small averaged stress enhances or sensitizes system response, while large average stress is an overwhelming, meaningless noise which inhibits the system, causing a blunted HPA response. Autoimmune disease becomes, then, an interaction between reset-to-zero cognition and HPA axis cognition, as modulated by the signal of applied external stress, which writes an image of itself on both initiation and promotion of disease.

Although certainly a result of gene-environment-development interaction – the inevitable 'triple helix' in the sense of Lewontin (2000) – rising rates and seriousness of SLE among African-American women over the last thirty years obviate simple genetic explanations: the genetic structure of the US population has not changed suddenly. What has changed, however, is the environmental and developmental context for African-American women. In particular, since the end of World War II processes of 'urban renewal', and after about 1970, of policy-driven contagious urban decay, together amounting to massive urban desertification, have left vast tracts of what had been thriving African-American urban neighborhoods looking like Dresden after the firebombing (e.g. D. Wallace and R. Wallace, 1998; R. Wallace and D. Wallace, 1997). African-American women born since the middle 1970's have not had the stable childhood developmental environment of their predecessors. African-American women born before the 1970's have not had a stable aging environment, affecting another developmental process.

The contagious 'hollowing out' of African-American neighborhoods in the majority of major US cities – New York, Philadelphia, Cleveland, Toledo, Detroit, St. Louis, and so on – has been compounded by, and intertwined with, a massive deindustrialization driven by the aftermath of the Cold War which Ullmann (1988), Melman (1971) and others claim is a consequence of the massive diversion of scientific and engineering resources from civilian to military enterprise during the Cold War. They argue that a healthy US manufacturing economy required at least a 3 percent annual improvement in productivity to maintain itself against foreign competition, a rate of growth fueled by our considerable technical resources through about 1965. Thereafter, increasing consumption of scientific and engineering personnel by the military and aerospace enterprises of the Cold War, and the consequent shift in technological emphasis away from the needs of the civilian economy, made this rate of improvement impossible, and rapid US deindustrialization ensued. Wallace et al. (1997, 1999) and Wallace and Wallace (1997) examine the consequences of this collapse for the US AIDS epidemic and other public health problems.

Deindustrialization, like policy-driven contagious urban decay and 'urban renewal' which together have hollowed out most African-American urban neighborhoods, represents the permanent dispersal of social, economic, and political capital within the worst affected regions, particularly as deindustrialization is now convoluted with urban decay.

Pappas, (1989) describes the general circumstances:

> "By 1982 mass unemployment had reemerged as a major social issue [in the USA]. Unemployment rose to its highest level since before World War II, and an estimated 12 million people were out of work – 10.8 percent of the labor force in the nation. It was not, however, a really new phenomenon. After 1968 a pattern was established in which each recession was followed by higher levels of unemployment during recovery. During the depth of the 1975 recession, national unemployment rose to 9.2 percent. In 1983, when a recovery was proclaimed, unemployment remained at 9.5 percent annually.
>
> Certain sectors of the work force have been more heavily affected than others. There was a 16.9 percent jobless rate among blue-collar workers in April, 1982... Unemployment and underemployment have become major problems for the working class. While monthly unemployment figures rise and fall, these underlying problems have persisted over a long period. Mild recoveries merely distract out attention from them."

Massey (1990) explores the particularly acute effect of this phenomenon on the African-American minority:

> "The decline of manufacturing, the suburbanization of blue-collar employment, and the rise of the service sector eliminated many well-paying jobs for unskilled minorities and reduced the pool of marriageable men, thereby undermining the strength of the family, increasing the rate of poverty, and isolating many inner-city residents from accessible, middle-class occupations."

It is the massive social disintegration of the last thirty years – particularly the widespread destruction of urban minority communities – which we believe is literally writing an image of itself upon African-American women as rising rates of SLE, among other things. Making that argument requires some further development. Before proceeding, however, some comment on methodology is required.

We adapt recent advances in understanding 'punctuated equilibrium' in evolutionary process (e.g. Wallace, 2002b; Wallace and Wallace, 1998, 1999; Wallace et al., 2003) to the question of how embedding structured psychosocial stress affects the interaction of 'mind' and 'body', and specifically seek to determine how the synergism of such stress with cognitive submodule function might be constrained by certain of the asymptotic limit theorems of probability.

We know that, regardless of the probability distribution of a particular stochastic variate, the Central Limit Theorem en-sures that long sums of independent realizations of that variate will follow a Normal distribution. Analogous constraints exist on the behavior of the 'information sources' we will find associated with both structured stress and cognitive module function – both independent and interacting – and these are described by the limit theorems of information theory. Imposition of phase transition formalism from statistical physics, in the spirit of the Large Deviations Program of applied probability, permits concise and unified description of evolutionary and cognitive 'learning plateaus' which, in the evolutionary case, are interpreted as evolutionary punctuation (e.g. Wallace, 2002a, b). This approach provides a 'natural' means of exploring punctuated processes in the effects of structured stress on mind-body interaction, in this case related to SLE and other autoimmune disorders.

The model, as in the relation of the Central Limit Theorem to parametric statistical inference, is almost independent of the detailed structure of the interacting information sources inevitably associated with cognitive process, important as such structure may be in other contexts. This finesses some of the profound ambiguities associated with 'dynamic systems theory' and 'deterministic chaos' treatments in which the existence of 'dynamic attractors' depends on very specific kinds of differential equation models akin to those used to describe ecological population dynamics, chemical processes, or physical systems of weights-on-springs. Cognitive phenomena are neither well-stirred Erlenmeyer flasks of reacting agents, nor distorted mechanical clocks. Indeed, much of contemporary nonlinear dynamics can be subsumed within our formalism through 'symbolic dynamics' discretization techniques (e.g. Beck and Schlogl, 1995).

Rather than taking symbolic dynamics as an approximation to 'more exact' nonlinear ordinary or stochastic differential equation models, we throw out, as it were, the Cheshire cat and keep the cat's smile, generalizing symbolic dynamics to a more comprehensive information dynamics not constrained by 18th Century ghosts trapped in noisy, nonlinear, mechanical devices.

Our approach is conditioned by Waddington's (1972) vision that, in situations which arise when there is mutual interaction between the complexity-out-of-simplicity (self-assembly), and simplicity-out-of-complexity (self-organization), processes are to be discussed most profoundly with the help of the analogy of language, i.e. that *language* may become a paradigm go a Theory of General Biology, but a language in which basic sentences are programs, not statements.

In contrast to nonlinear systems theory approaches in which it appears impossible to actually write down the assumed underlying 'basic nonlinear equations' of cognitive phenomena, it does seem possible to uncover the grammar and syntax of both structured psychosocial stress and the function of cognitive submodules, and to express their relations in terms of empirically observed regression models relating measurable biomarkers, behaviors, beliefs, feelings, and so on.

Our analysis will focus on the eigenstructure of those models, constrained by the behavior of information sources under appropriate asymptotic limit theorems of probability.

Clearly, then, our approach takes much from parametric statistics, and, while idiosyncratic 'nonparametric' models may be required in special cases, we may well capture the essence of the most common relevant phenomena.

## The model

**1. Cognition as language.** Atlan and Cohen (1998) and Cohen (2000) argue that the essence of immune cognition is comparison of a perceived antigenic signal with an internal, learned picture of the world, and then, upon that comparison, the choice of one response from a large repertoire of possible responses. Following the approach of Wallace (2000, 2002a), we make a very general, model of that process.

Pattern recognition-and-response, as we characterize it, proceeds by convoluting (i.e. comparing) an incoming external 'sensory' antigenic signal with an internal 'ongoing activity' – the 'learned picture of the world' – and, at some point, triggering an appropriate action based on a decision that the pattern of sensory activity requires a response. We need not model how the pattern recognition system is 'trained', and hence we adopt a weak model, regardless of learning paradigm, which can itself be more formally described by the Rate Distortion Theorem. We will, fulfilling Atlan and Cohen's (1998) criterion of meaning-from-response, define a language's contextual meaning entirely in terms of system output.

The model is as follows.

A pattern of sensory input is convoluted (compared) with internal 'ongoing' activity to create a path of convoluted signal $x = (a_0, a_1, ..., a_n, ...)$. This path is fed into a highly nonlinear 'decision oscillator' which generates an output $h(x)$ that is an element of one of two (presumably) disjoint sets $B_0$ and $B_1$. We take

$$B_0 \equiv b_0, ..., b_k,$$

$$B_1 \equiv b_{k+1}, ..., b_m.$$

Thus we permit a graded response, supposing that if

$$h(x) \in B_0$$

the pattern is not recognized, and if

$$h(x) \in B_1$$

the pattern is recognized and some action $b_j, k+1 \leq j \leq m$ takes place.

We are interested in paths $x$ which trigger pattern recognition-and-response exactly once. That is, given a fixed initial state $a_0$, such that $h(a_0) \in B_0$, we examine all possible subsequent paths $x$ beginning with $a_0$ and leading exactly once to the event $h(x) \in B_1$. Thus $h(a_0, ..., a_j) \in B_0$ for all $j < m$, but $h(a_0, ..., a_m) \in B_1$.

For each positive integer $n$ let $N(n)$ be the number of paths of length $n$ which begin with some particular $a_0$ having $h(a_0) \in B_0$ and lead to the condition $h(x) \in B_1$. We shall call such paths 'meaningful' and assume $N(n)$ to be considerably less than the number of all possible paths of length $n$ – pattern recognition-and-response is comparatively rare. We further assume that the finite limit

$$H \equiv \lim_{n \to \infty} \frac{\log[N(n)]}{n}$$

both exists and is independent of the path $x$. We will – not surprisingly – call such a pattern recognition-and-response cognitive process *ergodic*.

We may thus define an ergodic information source $\mathbf{X}$ associated with stochastic variates $X_j$ having joint and conditional probabilities $P(a_0, ..., a_n)$ and $P(a_n|a_0, ..., a_{n-1})$ such that appropriate cross-sectional joint and conditional Shannon uncertainties may be defined which satisfy the relations

$$H[\mathbf{X}] = \lim_{n \to \infty} \frac{\log[N(n)]}{n} =$$

$$\lim_{n \to \infty} H(X_n|X_0, ..., X_{n-1}) =$$

$$\lim_{n \to \infty} \frac{H(X_0, ..., X_n)}{n}.$$

(1)

This statement is the essence of the Shannon-McMillan Theorem. The 'splitting criteria' $H(...)$ are all of the form

$$H = -\sum_{j=1}^{m} P_j \log[P_j],$$

where the $P_j$ are appropriate individual, joint, or conditional probabilities, so that $\sum_j P_j = 1$. Sometimes we substitute a 'mutual information' made up of sums of $H$-terms as the splitting criterion, as in the Rate Distortion and Joint Asymptotic Equipartition Theorems (RDT, JAEPT).

We say the inferred information source $\mathbf{X}$ is *dual* to the ergodic cognitive process, implicitly invoking a generalized 'language-of-thought' formalism which is not 'linguistically complete' in the sense of Chomsky (e.g Fodor, 1975).

Different 'languages' will, of course, be defined by different divisions of the total universe of possible responses into different pairs of sets $B_0$ and $B_1$, or by requiring more than one response in $B_1$ along a path. Like the use of different distortion measures in the Rate Distortion Theorem (e.g. Cover and Thomas, 1991), however, it seems obvious that the underlying dynamics will all be qualitatively similar. Dividing the full set of possible responses into the sets $B_0$ and $B_1$ may itself require 'higher order' cognitive decisions by other modules, suggesting the necessity of 'choice' within a more or less broad set of possible 'languages of thought'. This would directly reflect the need to 'shift gears' according to the different challenges faced by the organism, either cross-sectionally at a particular time, or developmentally as it matures, accounting for 'critical periods' in the onset of developmental disorder. A central problem then becomes the choice of a 'normal' zero-mode language among a very large set of possible languages representing the (hyper- or hypo-) 'excited states' accessible to the system. This is a fundamental point which we explore below in various ways.

Meaningful paths – creating an inherent grammar and syntax – are defined entirely in terms of system response, as Atlan and Cohen (1998) propose.

We can apply this formalism to the stochastic neuron in a neural network: A series of inputs $y_i^j, i = 1, ...m$ from $m$ nearby neurons at time $j$ to the neuron of interest is convoluted with 'weights' $w_i^j, i = 1, ..., m$, using an inner product

$$a_j = \mathbf{y}^j \cdot \mathbf{w}^j \equiv \sum_{i=1}^{m} y_i^j w_i^j$$

(2)

in the context of a 'transfer function' $f(\mathbf{y}^j \cdot \mathbf{w}^j)$ such that the probability of the neuron firing and having a discrete output $z^j = 1$ is $P(z^j = 1) = f(\mathbf{y}^j \cdot \mathbf{w}^j)$.

Thus the probability that the neuron does not fire at time j is just $1 - P$. In the usual terminology the $m$ values $y_i^j$ constitute the 'sensory activity' and the $m$ weights $w_i^j$ the 'ongoing activity' at time $j$, with $a_j = \mathbf{y}^j \cdot \mathbf{w}^j$ and the path $x \equiv a_0, a_1, ..., a_n, ...$. A little more work leads to a standard neural network model in which the network is trained by appropriately varying $\mathbf{w}$ through least squares or other error minimization feedback. This can be shown to replicate rate distortion arguments, as we can use the error definition to define a distortion function which measures the difference between the training pattern $y$ and the network output $\hat{y}$ as a function, for example, of the inverse number of training cycles, $K$. As we will discuss in another context, 'learning plateau' behavior emerges naturally as a phase transition in the parameter $K$ in the mutual information $I(Y, \hat{Y})$.

Thus we will eventually parametize the information source uncertainty of the dual information source to a cognitive pattern recognition-and-response with respect to one or more variates, writing, e.g. $H[\mathbf{K}]$, where $\mathbf{K} \equiv (K_1, ..., K_s)$ represents a vector in a parameter space. Let the vector $\mathbf{K}$ follow some path in time, i.e. trace out a generalized line or surface $\mathbf{K}(t)$. We will, following the argument of Wallace (2002b), assume that the probabilities defining $H$, for the most part, closely track changes in $\mathbf{K}(t)$, so that along a particular 'piece' of a path in parameter space the information source remains as close to memoryless and ergodic as is needed for the mathematics to work. Between pieces we impose phase transition characterized by a renormalization symmetry, in the sense of Wilson (1971). See Binney, et al. (1986) for a more complete discussion.

We will call such an information source 'adiabatically piecewise memoryless ergodic' (APME). The ergodic nature of the information sources is a generalization of the 'law of large numbers' and implies that the long-time averages we will need to calculate can, in fact, be closely approximated by averages across the probability spaces of those sources. This is no small matter.

**2. Nonorthogonal eigenmodes of the immunological homunculus.** Cohen (2000) defines the immunological homunculus as the immune system's image of the self, i.e.

its characteristic pattern of response to self-antigens through autoantibodies, as regulated by a set of anti-autoantibodies. Other models are possible, e.g. the thymus conditioning approach of Whittingham and Mackey (1971). Whatever model used, we assume the immune system can monitor the self through examination of fluctuations in the immunological homunculus, and respond accordingly in a coherent, programmatic manner.

Infection, chemical exposure, stress, or physical injury, will perturb the immune system's image of the self, which, after successful address, will return to the 'zero order state'. To anticipate the argument, an essential problem for the immune system is to recognize what that state is, among all possible pictures of the self represented by the possible states of the immune homunculus: There is no 'energy minimization' strategy which permits a simple identification of such a state, since all states of the dynamic immunological homunculus are, effectively, high energy states. We argue that such recognition requires, in fact, a fairly sophisticated 'second order' cognitive decision. That is, identifying a baseline states is itself a cognitive process.

We are arguing, essentially, for the immunological equivalent of the 'mental modules' which evolutionary psychologists argue must have evolved in the human mind to efficiently process particular sets of sensory data, especially those related to human social interactions – recognition of facial expression, language, perhaps pheromone detection (e.g. Barkow et al. , 1992). Elsewhere (Wallace et al., 2003), following Nunney (1999), we have already argued that the body's 'tumor control strategy' must, in effect, be such a cognitive module, made increasingly complicated for large animals by the interaction of cell number with organism longevity. Here we extend that argument to autoimmunity.

We assume that the immune homunculus can be represented by some elaborate system of nonlinear equations, possibly involving cytokine, self-antibody, and anti-self-antibody concentrations and their spatial distributions. Our principal assumption is that the 'zero order state' is learned, and that changes about that state induced by external perturbations are relatively small.

We further assume that, expanding about the 'reference state', all variables, $x_i$ depend on all others 'nearly linearly', so that we can write, to first order at time $t$, a system of empirical regression equations describing the cascade of cytokines:

$$x_i(t) = \sum_{j \neq i}^{s} b_{i,j} x_j(t) + \epsilon_i(t, x_1(t), ..., x_s(t)).$$

(3)

At reference, the $x_i$ are defined to be zero, as are the $\epsilon_i$. Most critically, we will assume the $b_{i,j}$ *have been determined from empirical regression relations*. This assumption provides the mathematical foundation for our analysis.

Here the $x_j, j = 1, ..., s$ are both 'independent' and 'dependent' variables involved in the inevitable cytokine feedback cascade about the 'reference configuration', and the $\epsilon_i$ represent 'error terms' which are not necessarily small in this approximation. $s$ may be fairly large, depending, presumably, on the size of the animal according to some power law. Note that the $\epsilon$ terms will become external perturbations in the subsequent analysis.

In matrix notation this set of equations becomes

$$X(t) = \mathbf{B}X(t) + U(t)$$

(4)

where $X(t)$ is an $s$-dimensional vector, $\mathbf{B}$ is an $s \times s$ matrix of regression coefficients having a zero diagonal, and $U$ is an $s$-dimensional vector containing 'error' terms which are not necessarily small. We suggest that, on the timescale of applied perturbations and initial responses, the $\mathbf{B}$-matrix remains relatively constant.

This structure, by virtue of its determination through least squares linear regression, has a number of interesting properties which permit estimation of the effects of a perturbation. Rewriting, we obtain

$$[\mathbf{I} - \mathbf{B}]X(t) = U(t)$$

(5)

where $\mathbf{I}$ is the $s \times s$ identity matrix and, to reiterate, $\mathbf{B}$ has a zero diagonal.

We next *reexpress matters in terms of the eigenstructure of* $\mathbf{B}$.

Let $\mathbf{Q}$ be the matrix of eigenvectors which diagonalizes $\mathbf{B}$ (or at least reduces it to block-diagonal Jordan canonical form). Take $\mathbf{Q}Y(t) = X(t)$ and $\mathbf{Q}W(t) = U(t)$. Let $\mathbf{J}$ be the diagonal matrix of eigenvalues of $\mathbf{B}$ so that $\mathbf{B} = \mathbf{QJQ}^{-1}$. The eigenvalues of $\mathbf{B}$ can be shown to all be real (D. Wallace and R. Wallace, 2000). Then, for the eigenvectors $Y_k$ of $\mathbf{B}$, corresponding to eigenvalues $\lambda_k$,

$$Y_k(t) = \mathbf{J}Y_k(t) + W_k(t).$$

(6)

Using a term-by-term shorthand for the components of $Y_k$, this becomes

$$y_k(t) = \lambda_k y_k(t) + w_k(t).$$

Define the mean of a time dependent function $f(t)$ over the time interval $[0, \Delta T]$ as

$$E[f] \equiv \frac{1}{\Delta T} \int_0^{\Delta T} f(t)dt.$$

(7)

We assume an appropriately rational structure as $\Delta T \to \infty$; in particular we can, if necessary, take the perturbing signals to be the output of an ergodic information source, and in general require that all of the resulting signals will reflect that structure, so that time averages can generally be replaced by ensemble averages on the appropriate timescales.

The variance $V[f]$ over the same time interval is defined as $E[(f - E[f])^2]$. Taking matters again term-by-term, we obtain

$$V[(1-\lambda_k)y_k] = V[w_k]$$

so that

$$V[y_k] = \frac{V[w_k]}{(1-\lambda_k)^2}$$

or

$$\sigma(y_k) = \frac{\sigma(w_k)}{|1-\lambda_k|},$$

(8)

where $\sigma$ represents the standard deviation.

The $y_k$ are the components of the eigentransformed immune system variates, and the $w_k$ are the similarly transformed variates of the driving externalities of infection, injury, or stress, as perceived by the immune homunculus. These signals are not, in particular, always likely to be random, and may have highly structured internal serial correlations of 'grammar' and 'syntax', although this assumption is not necessary here.

The eigenvectors $Y_k$ are characteristic but non-orthogonal combinations of the original variates $X_k$ whose standard deviation is that of the transformed externalities *amplified by the term* $1/|1-\lambda|$. Characteristic patterns of perturbation $w$ can therefore trigger characteristic, but nonorthogonal, amplified patterns of general response $Y_k$ in the immune homunculus, which must, among other things, instantiate the immune system's perception of the self. Although there may be $s$ of these 'excited eigenmodes' – in addition to the zero reference state – relatively few of them will be highly probable. It is these highly probable $Y_k$ which we propose form the possible set of defined zero states of the immune homunculus, one of which, including the initial '$Y_0 = 0$' state, must be chosen by a cognitive reset-to-zero module.

Note in particular that the nonorthogonal nature of the eigenstructure of **B** implies plieotropy, i.e. that a single input signal may have multiple possible outputs, here in proportion to the magnitude of the excitation. Such plieotropy, in this context, suggests the possibility of comorbidity in autoimmune diseases.

Extension of the model to include rates of change of cytokine concentrations and the like, in addition to their magnitudes, is algebraically complicated but seems fairly direct. Intuitively, such extension must give a first order matrix relation much like equation (4), but now in terms of both the $x_j$ and their time rates of change $\dot{x}_j$. Such matrix equations typically have eigenmodes with *complex* eigenvalues representing linked patterns of dynamical limit cycles, rather than simple fixed eigenstructures. Thus the problem becomes one of perturbation from a reference pattern of limit cycles, and of characterizing the behavior of a cognitive submodule permitting identification of that reference pattern.

**3. Circadian and other cycles.** If the immunological homunculus could be *entirely* described by a simple system of first order differential equations, expanded near a zero-state, then under perturbation we would have something like

$$\dot{X}(t) = \mathbf{R}X(t) + \epsilon(t)$$

(9)

where $X(t)$ is the vector of displacements from the zero state, $\epsilon(t)$ the vector of perturbations, and **R** an appropriate fixed matrix of real numbers, here having purely imaginary eigenvalues. Thus the trace of **R**, the sum of the real parts of the eigenvalues, is zero, analogous to the condition on the $B$-matrix above. This is a simple version of the famous Langevin equation, with a full solution in terms of the Fokker-Planck equation if $\epsilon(t)$ has appropriately random 'white noise' properties, an analysis which moves rapidly into the realm of stochastic differential equations. Unfortunately, $\epsilon(t)$ is unlikely to be random in the sense necessary for such an approach, which requires the covariance of $\epsilon$ between different times to be proportional to a delta function. We would, on the contrary, generally expect $\epsilon(t)$ to be the output of an appropriately regular information source, with elaborate covariance structure.

We can, however, make a simplified treatment quite like the previous development, provided the system is asymptotically bounded in the sense that, for all time $t$, there is a fixed, positive real number $c$ such that, for all components of the vector $X$,

$$|x_j(t)| \leq c.$$

(10)

That is, the system cannot move arbitrarily far from its 'zero state'. Then, taking the time average of equation (9) in the sense of the previous development gives

$$E[\dot{X}] = \mathbf{R}E[X] + E[\epsilon], \tag{11}$$

where we very explicitly do not assume the perturbations are random with zero mean. Implicitly, then, we are assigning a grammar and syntax to the perturbing structures, and, if the associated information source is ergodic, we can replace the time averages with ensemble averages, across the probability space of the 'language' of perturbation.

Writing out the left hand side of the equation gives, component by component

$$E[\dot{x}_j] = \lim_{\Delta T \to \infty} \frac{1}{\Delta T} \int_0^{\Delta T} [dx_j/dt]dt$$

$$= \lim_{\Delta T \to \infty} \frac{1}{\Delta T}[x_j(\Delta T) - x_j(0)].$$

This expression is bounded both above and below by

$$\lim_{\Delta T \to \infty} \frac{2c}{\Delta T} = 0,$$

and is thus itself zero.

We obtain, then,

$$\mathbf{R}E[X] = -E[\epsilon]. \tag{12}$$

Somewhat heuristically, to first order an eigentransformation in terms of $\mathbf{R}$ gives a result analogous to equation (8), again component-by-component:

$$E[y_k] = -\frac{E[w_k]}{\omega_k}. \tag{13}$$

$Y_k$ is the k-th eigenvector of $\mathbf{R}$, $W_k$ is the *eigentransformed* perturbation, and $\omega_k$ is the frequency of the cyclic eigenmode $Y_k$.

The equation states that cyclic eigenmode amplification by nonrandom structured external perturbation is inversely proportional to eigenmode frequency. That is, slower cycles are amplified by perturbation more than rapid ones, in proportion to their period. Again, there is no orthogonality constraint on the eigenvectors of $\mathbf{R}$, suggesting the possibility of plieotropic response.

This is an interesting result: Many physiological cycles are characterized by several relatively slow processes, in comparison with the 'standard physiological clock' of the heartbeat: daily circadian, monthly hormonal, and annual light/temperature cycles. Pathologically amplified (nonorthogonal) eigenmodes – displacements from zero related to autoimmune disease – according to this argument, may well be intimately associated with these cycles. The monthly hormonal cycle of non-menopausal women might then be related to a particular form of 'non-zero offset', i.e. an excited mode representing a particular autoimmune disease. Similarly, tropical populations could suffer less from excited modes associated with annual cycles of light and temperature (and their ecological sequelae), perhaps accounting for the 'tropical gradient' in multiple sclerosis, an autoimmune disease of the central nervous system. That is, autoimmune disease might well be classifiable by associated cycle or cycles, as well as by perturbation-of-onset.

Most autoimmune diseases would seem, of necessity, to be particularly related to the circadian cycle, which is universal, very powerful, and always fairly long compared to the heartbeat. Thus autoimmune diseases may, from this development, be especially stratified by their disturbance in various circadian rhythms (e.g. Lechner et al., 2000; Hilty et al., 2000).

Schubert et al. (1999) report a particularly interesting experiment involving a long time series of daily monitoring of urine neopterin in a white European patient with SLE. The particular focus was the association of concentration spikes with the grammar and syntax of life stressors. Urine neopterin above 300 $\mu$-mol is predictive of SLE activity. Not all daily stressors, some quite extreme, caused significant spiking. For this patient, a raised neopterin level was triggered only by incidents of unusual conflict with close family members: thus only a very few characteristic stresses were 'meaningful' in the context of SLE, most others were not. This work strongly suggests the validity of an 'information' approach.

**4. Plieotropy: the retina of the immunological homunculus.** A slight variation of the model above leads to further interesting speculations. Rather than taking a differential equation approach, we follow the suggestion of Schubert et al. (1999) and suppose that the daily circadian or some other cycle imposes a periodic temporal structure on the immunological homunculus, and we measure in units of that period in the sense that the 'state' of the homunculus at some time $t + 1$, which we write $X_{t+1}$, is assumed to be a function of its state at time $t$:

$$X_{t+1} = \mathbf{R}_{t+1}X_t. \tag{14}$$

If $X_t$, the simplified, internal picture of the body at time $t$, is of dimension $m$, then $\mathbf{R}_t$, the manner in which that picture changes in time (from time $t$ to $t+1$), has $m^2$ components. If the state of the homunculus at time $t = 0$ is $X_0$, then iterating the relation above gives the state at time $t$ as

$$X_t = \mathbf{R}_t \mathbf{R}_{t-1} \mathbf{R}_{t-2} ... \mathbf{R}_1 X_0.$$

(15)

The state of the body is, in this picture, essentially represented by an information-theoretic path defined by the stochastic sequence in $\mathbf{R}_t$, each member having $m^2$ components: the grammar and syntax of how things change tells us much about how we are. That sequence is mapped onto a parallel path in the states of the immunological homunculus, the set $X_0, X_1, ..., X_t$, each having $m$ components.

If the state of the body can, in fact, be characterized as an information source – a generalized language – so that the paths of $\mathbf{R}_t$ are autocorrelated, then the autocorrelated paths in $X(t)$ represent the output of a parallel information source which is, Rate Distortion arguments to the contrary, apparently a greatly simplified, and thus grossly distorted, picture of that body.

This may not necessarily be the case.

Let us examine a single iteration in more detail, assuming now that there is a zero reference state, $\mathbf{R}_0$, for the sequence in $\mathbf{R}_t$, and that

$$X_{t+1} = (\mathbf{R}_0 + \delta \mathbf{R}_{t+1}) X_t,$$

(16)

where $\delta \mathbf{R}_t$ is 'small' in some sense compared to $\mathbf{R}_0$.

We again invoke a diagonalization in terms of $\mathbf{R}_0$. Let $\mathbf{Q}$ be the matrix of eigenvectors which (Jordan) diagonalizes $\mathbf{R}_0$. Then we can write

$$\mathbf{Q} X_{t+1} = (\mathbf{Q} \mathbf{R}_0 \mathbf{Q}^{-1} + \mathbf{Q} \delta \mathbf{R}_{t+1} \mathbf{Q}^{-1}) \mathbf{Q} X_t.$$

If we take $\mathbf{Q} X_t$ to be an eigenvector of $\mathbf{R}_0$, say $Y_k$, with eigenvalue $\lambda_k$, we can rewrite this equation as a spectral expansion,

$$Y_{t+1} = (\mathbf{J} + \delta \mathbf{J}_{t+1}) Y_k \equiv \lambda_k Y_k + \delta Y_{t+1} =$$

$$\lambda_k Y_k + \sum_{j=1}^{n} a_j Y_j,$$

(17)

where $\mathbf{J}$ is a (block) diagonal matrix, $\delta \mathbf{J}_{t+1} \equiv \mathbf{Q} \mathbf{R}_{t+1} \mathbf{Q}^{-1}$, and $\delta Y_{t+1}$ *has been expanded in terms of a spectrum of the eigenvectors of* $\mathbf{R}_0$, with

$$|a_j| \ll |\lambda_k|, |a_{j+1}| \ll |a_j|.$$

(18)

The essential point is that, provided $\mathbf{R}_0$ is chosen or 'tuned' so that this condition is true, the first few terms in the spectrum of the plieotropic iteration of the eigenstate will contain most of the essential information about the perturbation. We envision this as similar to the detection of color in the retina, where three overlapping non-orthogonal 'eigenmodes' of response suffice to characterize a vast plethora of color sensation. Here, if such a spectral analysis is possible, a very small number of eigenmodes of the immunological homunculus would suffice to permit identification of a vast range of perturbed bodily states: the rate-distortion constraints become very manageable indeed.

This is a necessarily more complex process than color detection since the immune system has both innate and learned components, and genetic programming is of limited value. The key to the problem, we believe, would lie in the proper rate-distortion tuning of the system, i.e. the choice of zero-mode, $\mathbf{R}_0$. Such choice, we suspect, is very complicated and requires a dedicated cognitive submodule of immune cognition, one whose disturbance itself represents onset of an autoimmune disease.

Caswell (2001) provides an accessible introduction to the kind of matrix population models we have invoked.

**5. Circadian-hormonal cycle synergism.** While men and women share both circadian and annual cycles, women of reproductive age are likely to find any daily physiological patterns interacting with their hormonal cycle – stereotypically 28 days in length. This, we will show, potentially creates a vast complication for any reset-to-zero cognitive module, and may significantly contribute to the comparatively higher rate of autoimmune disease among women than men in 'Westernized' countries. The argument is quite direct.

We take the female hormonal cycle to be made up of $m$ circadian cycles, so that we can, over a single such cycle, write

$$X_{t+m} = \mathbf{R}_m \mathbf{R}_{m-1} ... \mathbf{R}_1 X_t$$

$$= \mathbf{Q}_1 X_t,$$

(19)

with

$$\mathbf{Q}_1 \equiv \mathbf{R}_m ... \mathbf{R}_1.$$

We are, in effect, doing matrix algebra 'modulo $m$'.

Without going into details of a 'modulo $m$' expansion of equation (17) in terms of the components of the product matrix $\mathbf{Q}$, it is evident that the question of the phase of the product expansion is of considerable importance. That is, given a cyclic expansion for the $\mathbf{Q}$, where does one start? Which day is the first day? The matricies $\mathbf{R}_h$ may differ significantly among themselves, depending on daily perturbations. In particular, as Caswell (2001, p. 351) discusses and equation (17) implies, the eigenvectors of the $\mathbf{R}_h$ are all different. Thus the eigenstructure of the composite $\mathbf{Q}$, and hence the definition of the 'zero state' such that $\mathbf{Q}_t = \mathbf{Q}_0 + \delta\mathbf{Q}_t$, depends on where one begins in the cycle, i.e. the phase. There are, then, $m$ possible choices for a $\mathbf{Q}_0$.

For women, any reset-to-zero immunological module must choose not only the appropriate daily $\mathbf{R}_0$, but choose among $m$ possible monthly $\mathbf{Q}_0$. This is a composite choice fraught with possibilities for error which a similar immune cognitive module in males would not confront. We suggest this complexity may indeed account for the relatively higher rates of certain autoimmune diseases among women.

**6. The cognitive HPA axis.** Atlan and Cohen (1998) argue that the essence of cognition is comparison of a perceived external signal with an internal, learned, picture of the world, and then upon that comparison, the choice of a response from a much larger repertory of possible responses. Clearly, from this perspective, the HPA axis, the 'flight-or-fight' reflex, is cognitive. Upon recognition of a new perturbation in the surrounding environment, memory and brain cognition evaluate and choose from several possible responses, (1) no action needed, (2) flight, (3) fight, (4) helplessness (i.e. flight or fight needed but not possible). Upon appropriate conditioning the HPA axis is able to accelerate the decision process, much like the immune system has a more efficient response to second pathogenic challenge once the initial infection has become encoded in immune memory. For example, 'hyperreactivity' in the context of PTSD is well known. We thus take the HPA axis to be cognitive in the sense of section 1 above, and thus associated with a dual information source having a 'grammar' and 'syntax', in a large sense.

Stress, as we envision it, is not a random sequence of perturbations, and is not independent of its perception. Rather, it involves a highly correlated, grammatical, syntactical process by which an embedding psychosocial environment communicates with an individual, particularly with that individual's HPA axis, in the context of social hierarchy. We view the stress experienced by an individual as APME information source, interacting with a similar dual information source defined by HPA axis cognition.

Again, the ergodic nature of the 'language' of stress is essentially a generalization of the law of large numbers, so that long-time averages can be well approximated by cross-sectional expectations. Languages do not have simple autocorrelation patterns, in distinct contrast with the usual assumption of random perturbations by 'white noise' in the standard formulation of stochastic differential equations.

Let us suppose we cannot measure stress, but can determine the concentrations of HPA axis hormones and other biochemicals according to some 'natural' time frame, which we will characterize as the inherent period of the system. Suppose, in the absence of extraordinary 'meaningful' psychosocial stress, we measure a series of $n$ concentrations at time $t$ which we represent as an $n$-dimensional vector $X_t$. Suppose we conduct a number of experiments, and create a regression model so that we can, in the absence of perturbation, write, to first order, the concentration of biomarkers at time $t+1$ in terms of that at time $t$ using a matrix equation of the form

$$X_{t+1} \approx <\mathbf{R}> X_t + b_0, \tag{20}$$

where $<\mathbf{R}>$ is the matrix of regression coefficients and $b_0$ a vector of constant terms.

We then suppose that, in the presence of a perturbation by structured stress

$$X_{t+1} = (<\mathbf{R}> + \delta\mathbf{R}_{t+1})X_t + b_0$$

$$\equiv <\mathbf{R}> X_t + \epsilon_{t+1}, \tag{21}$$

where we have absorbed both $b_0$ and $\delta\mathbf{R}_{t+1} X_t$ into a vector $\epsilon_{t+1}$ of 'error' terms which are not necessarily small in this formulation. In addition it is important to realize that this is not a population process whose continuous analog is exponential growth. Rather what we examine is more akin to the passage of a signal – structured psychosocial stress – through a distorting physiological filter.

If the matrix of regression coefficients $<\mathbf{R}>$ is sufficiently regular, we can (Jordan block) diagonalize it using the matrix of its column eigenvectors $\mathbf{Q}$, writing

$$\mathbf{Q}X_{t+1} = (\mathbf{Q}<\mathbf{R}>\mathbf{Q}^{-1})\mathbf{Q}X_t + \mathbf{Q}\epsilon_{t+1}, \tag{22}$$

or equivalently as

$$Y_{t+1} = <\mathbf{J}> Y_t + W_{t+1}, \tag{23}$$

where $Y_t \equiv \mathbf{Q}X_t$, $W_{t+1} \equiv \mathbf{Q}\epsilon_{t+1}$, and $<\mathbf{J}> \equiv \mathbf{Q}<\mathbf{R}>\mathbf{Q}^{-1}$ is a (block) diagonal matrix in terms of the eigenvalues of $<\mathbf{R}>$.

Thus the (rate distorted) writing of structured stress on the HPA axis through $\delta \mathbf{R}_{t+1}$ is reexpressed in terms of the vector $W_{t+1}$.

The sequence of $W_{t+1}$ is the rate-distorted image of the information source defined by the system of external structured psychosocial stress. This formulation permits estimation of the long-term steady-state effects of that image on the HPA axis. The essential trick is to recognize that because everything is (APM) ergodic, we can either time or ensemble average both sides of equation (23), so that the one-period offset is absorbed in the averaging, giving an 'equilibrium' relation

$$<Y> = <\mathbf{J}><Y> + <W>$$

or

$$<Y> = (\mathbf{I} - <\mathbf{J}>)^{-1} <W>, \quad (24)$$

where $\mathbf{I}$ is the $n \times n$ identity matrix.

Now we reverse the argument: Suppose that $Y_k$ is chosen to be some fixed eigenvector of $<\mathbf{R}>$. Using the diagonalization of $<\mathbf{J}>$ in terms of its eigenvalues, we obtain the average excitation of the HPA axis in terms of some eigentransformed pattern of exciting perturbations as

$$<Y_k> = \frac{1}{1 - <\lambda_k>} <W_k> \quad (25)$$

where $<\lambda_k>$ is the eigenvalue of $<Y_k>$, and $<W_k>$ is some appropriately transformed set of ongoing perturbations by structured psychosocial stress.

The essence of this result is that *there will be a characteristic form of perturbation by structured psychosocial stress – the $W_k$ – which will resonantly excite a particular eigenmode of the HPA axis.* Conversely, by 'tuning' the eigenmodes of $<\mathbf{R}>$, the HPA axis can be trained to galvanized response in the presence of particular forms of perturbation.

This is because, if $<\mathbf{R}>$ has been appropriately determined from regression relations, then the $\lambda_k$ will be a kind of multiple correlation coefficient (e.g. Wallace and Wallace, 2000), so that particular eigenpatterns of perturbation will have greatly amplified impact on the behavior of the HPA axis.

If $\lambda = 0$ then perturbation has no more effect than its own magnitude. If, however, $\lambda \to 1$, then the written image of a perturbing psychosocial stressor will have very great effect on the HPA axis. Following Ives (1995), we call a system with $\lambda \approx 0$ *resilient* since its response is no greater than the perturbation itself.

We suggest, then, that learning by the HPA axis is, again, the process of tuning response to perturbation. This is why we have written $<\mathbf{R}>$ instead of simply $\mathbf{R}$: The regression matrix is a tunable set of variables. In turn, this suggests the possibility of another 'retina' argument like that leading to equation (17) above.

Suppose we require that $<\lambda>$ itself be a function of the magnitude of excitation, i.e.

$$<\lambda> = f(|<W>|)$$

where $|<W>|$ is the vector length of $<W>$. We can, for example, require the amplification factor $1/(1-<\lambda>)$ to have a signal transduction form, an inverted-U-shaped curve, for example the signal-to-noise ratio of a stochastic resonance, so that

$$\frac{1}{1 - <\lambda>} = \frac{1/|<W>|^2}{1 + b \exp[1/(2|<W>|)]}. \quad (26)$$

This places particular constraints on the behavior of the 'learned average' $<\mathbf{R}>$, and gives precisely the typical HPA axis pattern of initial hypersensitization, followed by anergy or 'burnout' with increasing average stress, a behavior that might well be characterized as 'pathological resilience', and may also have evolutionary significance.

Again variants of this model permit imposition of cycles of different length, for example hormonal on top of circadian. Typically this is done by requiring a cyclic structure in matrix multiplication, with a new matrix $\mathbf{S}$ defined in terms of a sequential set of the $\mathbf{R}$, having period $m$, so that

$$\mathbf{S}_t \equiv \mathbf{R}_{t+m} \mathbf{R}_{t+m-1} ... \mathbf{R}_t.$$

Again one does matrix algebra 'modulo m'.

To reiterate, while the eigenvalues of such a cyclic system may remain the same, its eigenvectors depend on the choice of phase, i.e. where you start in the cycle. This is a complexity of no small note, and could represent a source of contrast in HPA axis behavior between men and women, beyond that driven by the ten-fold difference in testosterone levels. Again, see Caswell (2001) for mathematical details.

Next we examine how the information source dual to the cognitive reset-to-zero process for fixed eigenmodes, eigenpatterns of limit cycles, or tuned spectra, can become linked in a punctuated manner with the cognitive HPA axis and a structured 'language' of external perturbation, a highly nontrivial development.

**7. Phase transitions of interacting information systems.** We suppose that the reset-to-zero cognitive module of the immune system devoted to the immunological homunculus, or the output of the cognitive HPA axis, can be represented by a sequence of 'states' in time, the 'path' $x \equiv x_0, x_1, ....$ Similarly, we assume an external signal of infection, tissue damage, chemical exposure, or 'psychosocial stress' can be similarly represented by a path $y \equiv y_0, y_1, ....$. These paths are, however, both very highly structured and, within themselves, are serially correlated and can, in fact, be represented by 'information sources' $\mathbf{X}$ and $\mathbf{Y}$. We assume the reset-to-zero cognitive process of the immunological homunculus and the external stressors interact, so that these sequences of states are not independent, but are jointly serially correlated. We can, then, define a path of sequential pairs as $z \equiv (x_0, y_0), (x_1, y_1), ....$.

The essential content of the Joint Asymptotic Equipartition Theorem is that the set of joint paths $z$ can be partitioned into a relatively small set of high probability which is termed *jointly typical*, and a much larger set of vanishingly small probability. Further, according to the JAEPT, the *splitting criterion* between high and low probability sets of pairs is the mutual information

$$I(X,Y) = H(X) - H(X|Y) = H(X) + H(Y) - H(X,Y)$$

(27)

where $H(X), H(Y), H(X|Y)$ and $H(X,Y)$ are, respectively, the Shannon uncertainties of $X$ and $Y$, their conditional uncertainty, and their joint uncertainty. See Cover and Thomas (1991) for mathematical details. Similar approaches to neural process have been recently adopted by Dimitrov and Miller (2001).

The high probability pairs of paths are, in this formulation, all equiprobable, and if $N(n)$ is the number of jointly typical pairs of length $n$, then

$$I(X,Y) = \lim_{n \to \infty} \frac{\log[N(n)]}{n}.$$

(28)

Generalizing the earlier language-on-a-network models of Wallace and Wallace (1998, 1999), we suppose there is a 'coupling parameter' $P$ representing the degree of linkage between the immune system's reset cognition and the system of external signals and stressors, and set $K = 1/P$, following the development of those earlier studies. Then we have

$$I[K] = \lim_{n \to \infty} \frac{\log[N(K,n)]}{n}.$$

The essential 'homology' between information theory and statistical mechanics lies in the similarity of this expression with the infinite volume limit of the free energy density. If $Z(K)$ is the statistical mechanics partition function derived from the system's Hamiltonian, then the free energy density is determined by the relation

$$F[K] = \lim_{V \to \infty} \frac{\log[Z(K)]}{V}.$$

(29)

$F$ is the free energy density, $V$ the system volume and $K = 1/T$, where $T$ is the system temperature.

We and others argue at some length (e.g. Wallace and Wallace, 1998, 1999; Rojdestvensky and Cottam, 2000; Feynman, 1996) that this is indeed a systematic mathematical homology which, we contend, permits importation of renormalization symmetry into information theory. Imposition of invariance under renormalization on the mutual information splitting criterion $I(X,Y)$ implies the existence of phase transitions analogous to learning plateaus or punctuated evolutionary equilibria in the relations between the cognitive reset mechanism and the system of external perturbations. A more complete mathematical treatment of these ideas is presented elsewhere, including detailed analysis of the way in which 'biological' phase transitions can formally differ from simple physical transitions (Wallace et al., 2003; Wallace and Wallace, 2003).

Elaborate developments are possible. From a the more limited perspective of the Rate Distortion Theorem we can view the onset of a punctuated interaction between the cognitive reset-to-zero mechanism of the immune homunculus and external stressors as a distorted image of the internal structure of those stressors within the homunculus:

Suppose that two (piecewise, adiabatically memoryless) ergodic information sources $\mathbf{Y}$ and $\mathbf{B}$ begin to interact, to 'talk' to each other, i.e. to influence each other in some way so that it is possible, for example, to look at the output of $\mathbf{B}$ – strings $b$ – and infer something about the behavior of $\mathbf{Y}$ from it – strings $y$. We suppose it possible to define a retranslation from the B-language into the Y-language through a deterministic code book, and call $\hat{\mathbf{Y}}$ the translated information source, as mirrored by $\mathbf{B}$.

Define some distortion measure comparing paths $y$ to paths $\hat{y}, d(y, \hat{y})$ (Cover and Thomas, 1991). We invoke the Rate Distortion Theorem's mutual information $I(Y, \hat{Y})$, which is the splitting criterion between high and low probability pairs of paths. Impose, now, a parametization by an inverse coupling strength $K$, and a renormalization symmetry representing the global structure of the system coupling.

Extending the analysis via *Network Information Theory*, triplets of sequences can be divided by a splitting criterion into two sets, having high and low probabilities respectively. For large $n$ the number of triplet sequences in the high probability set will be determined by the relation (Cover and Thomas, 1992, p. 387)

$$N(n) \propto \exp[nI(Y_1; Y_2|Y_3)],$$

(30)

where splitting criterion is given by

$$I(Y_1; Y_2|Y_3) \equiv$$

$$H(Y_3) + H(Y_1|Y_3) + H(Y_2|Y_3) - H(Y_1, Y_2, Y_3)$$

In this development $Y_3$ represents an embedding context for both $Y_1$ and $Y_2$.

We can then examine mixed cognitive/adaptive phase transitions analogous to learning plateaus (Wallace, 2002b) in the splitting criterion $I(Y_1, Y_2|Y_3)$. Note that our results are almost exactly parallel to the Eldredge/Gould model of evolutionary punctuated equilibrium (Eldredge, 1985; Gould, 2002).

If we wish to examine the impact of context, say the information source $Z$, on a larger number of interacting information sources representing cognitive modules, say the set $Y_j, j = 1..s$, we obtain a JAEPT splitting criterion of the form

$$I(Y_1, ..., Y_s|Z) = H(Z) + \sum_{j=1}^{s} H(Y_j|Z) - H(Y_1, ..., Y_s, Z).$$

### Autoimmune disease

According to current theory, the adapted human mind functions through the action and interaction of distinct mental modules which evolved fairly rapidly to help address special problems of environmental and social selection pressure faced by our Pleistocene ancestors (e.g. Barkow et al., 1992). As is well known in computer engineering, calculation by specialized submodules – e.g. numeric processor chips – can be a far more efficient means of solving particular well-defined classes of problems than direct computation by a generalized system. We suggest, then, that immune cognition has evolved specialized submodules to speed the address of certain commonly recurring challenges. Nunney (1999) has argued that, as a power law of cell count, specialized subsystems are increasingly required to recognize and redress tumorigenesis, mechanisms ranging from molecular error-correcting codes, to programmed cell death, and finally full-blown immune attack.

Here we argue that identification of the 'normal' state of the immunological homunculus – the immune system's self-image-module of the body – is a highly nontrivial task requiring a separate, specialized cognitive submodule within overall immune cognition. This is essentially because, for the vast majority of information systems, unlike a mechanical system, there are no 'restoring springs' whose low energy state automatically identifies equilibrium. That is, active comparison must be made of the state of the immunological homunculus with some stored internal reference picture, and a decision made about whether to reset to zero, a cognitive process. We further speculate that the complexity of such a submodule must also follow something like Nunney's power law with animal size, as the overall immune system and the immune image of the self, become increasingly complicated with rising number of cells.

Failure of that cognitive submodule results in identification of an 'excited' state of the immunological homunculus as 'normal', triggering the systematic patterns of self-antibody attack which constitute autoimmune disease, and which our analysis suggests may often be related to particular physiological cycles or signals which are long compared to heartbeat rate.

In sum, since such 'zero mode identification' (ZMI) is a (presumed) cognitive submodule of overall immune cognition, it involves a cognitive process, convoluting incoming 'sensory' with 'ongoing' internal memory data in choosing the zero state. The dual information source defined by this cognitive process can then interact in a punctuated manner with 'external information sources' according to the Rate Distortion arguments above. From a RDT perspective, then, those external information sources literally write a distorted image of themselves onto the ZMI in a punctuated manner: (relatively) sudden onset of an autoimmune disease whose progression is moderated by the HPA axis and its interactions with structured psychosocial stress.

Different systems of structured external signals – infections, chemical exposures, systems of 'psychosocial stress' – will, presumably, write different characteristic images of themselves onto the ZMI cognitive submodule, i.e. trigger different autoimmune diseases, perhaps stratified by their relation to circadian, hormonal, or annual cycles.

Zero mode identification is a more general problem for cognitive processes. For those dubious of the Homunculus regression model argument above, a brief abstract reformulation may be of interest. Recall that the essential characteristic of cognition in our formalism involves a function $h$ which maps a (convolutional) path $x = a_0, a_1, ..., a_n, ...$ onto a member of one of two disjoint sets, $B_0$ or $B_1$. Thus respectively, either (1) $h(x) \in B_0$, implying no action taken, or (2), $h(x) \in B_1$, and some particular response is chosen from a large repertoire of possible responses. We discussed briefly the problem of defining these two disjoint sets, and suggested that some 'higher order cognitive module' might be needed to identify what constituted $B_0$, the set of 'normal' states. Again, this is because there is no low energy mode for information systems: virtually all states are more or less high energy states, and there is no way to identify a ground state using the physicist's favorite variational or other minimization arguments on energy.

Suppose that higher order cognitive module, which we now recognize as a kind of Zero Mode Identification, interacts with an embedding language of structured psychosocial stress (or other systemic perturbation) and, instantiating a Rate Distortion image of that embedding stress, begins to include one or more members of the set $B_1$ into the set $B_0$. Recurrent 'hits' on that aberrant state would be experienced as episodes of pathology.

Empirical tests of this hypothesis, however, quickly lead again into real-world regression models involving the interrelations of measurable biomarkers, beliefs, behaviors, reported feelings, and so on, requiring formalism much like that used above.

The second stage of the argument – disease progression – involves interaction of the ZMI module with the HPA axis in the context of a chronic stress which can blunt HPA axis response.

The tool for this is an appropriate parametization of the network information theory mutual information $I(Y_1; Y_2|Y_3)$,

where $Y_1$ represents the dual information source of the ZMI module, $Y_2$ that of the HPA axis, and $Y_3$ the information source of the embedding 'language' of structured psychosocial stress, a cultural artifact affecting both $Y_1$ and $Y_2$.

Phase transition behavior of $I(Y_1;Y_2|Y_3)$ may be very complicated indeed. This suggests that the onset and staging of autoimmune disease, and their relation to the grammar and syntax of applied stressors, will be even more degenerate and plieotropic than is usual for strictly immune system phenomena. It would seem possible to apply yet another 'retina' argument, producing a limited spectrum of highly probable symptom patterns.

Iteration of this overall approach would permit introduction of even more linked cognitive submodules into a larger model, with concomitant increase in subtlety of behavior.

Further development would introduce the 'generalized Onsager relation' analysis of gradient effects in driving parameters which affects system behavior between phase transitions (e.g. Wallace, 2002a). All these extensions remain to be done, and are not trivial.

## Discussion and conclusions

Mathematical models of physiological and other ecosystems – like those we present here – are notorious for their unreliability, instability, and oversimplification. As it is said, "all models are wrong, but some models are useful". The mathematical ecologist E.C. Pielou (1977, p. 106) put the matter thus:

> "...[Mathematical] models are easy to devise; even though the assumptions of which they are constructed may be hard to justify, the magic phrase 'let us assume that...' overrides objections temporarily. One is then confronted with a much harder task: How is such a model to be tested? The correspondence between a model's predictions and observed events is sometimes gratifyingly close but this cannot be taken to imply the model's simplifying assumptions are reasonable in the sense that neglected complications are indeed negligible in their effects...
> In my opinion the usefulness of models is great... [however] it consists *not in answering questions but in raising them*. Models can be used to inspire new field investigations and these are the only source of new knowledge as opposed to new speculation."

Given this caveat, the speculations raised by our modeling exercise are of considerable interest.

Recent theories of coronary heart disease – CHD – (e.g. Ridker 2002; Libbey et al., 2002) identify a dynamic and progressive chronic vascular inflammation as the basic pathogenic biological mechanism, a process in which the cytokine IL-6 and C-reactive protein (CRP) play central roles. We have reviewed something of the "IL-6" hypothesis regarding the etiology SLE. An earlier analysis along these lines identified social structures of 'pathogenic social hierarchy' (PSH) in the US as critical in determining population-level patterns of CHD among African-American males (Wallace et al., 2002b). In that paper, historical cultural patterns of racism and discrimination were viewed as directly writing themselves onto the 'language' of immune cognition in a punctuated Rate Distortion manner to produce chronic vascular inflammation among subordinate populations.

Female hormones are known to be generally protective against CHD. Where, then, does the stress of PSH express itself in women? Figure 1 is taken from material on health and hierarchy in Singh-Manoux et al. (2003). It displays, for men and women separately, self-reported health as a function of self-reported status rank, where 1 is high and 10 low rank, among some 7,000 male and 3,400 female London-based office staff, aged 35-55 working in 20 Civil Service departments in the late 1990's. Self-reported health is a highly significant predictor of future morbidity and mortality.

Remarkably, the results for men and women are virtually indistinguishable in what is clearly a kind of toxicological dose-response curve, displaying physiological response against a 'dosage' of hierarchy which may include measures of both stress and real availability of resources (Link and Phelan, 2000).

We propose that PSH can also write an image of itself onto both a cognitive physiological module of the immune system, what we have called 'zero mode identification' which defines the 'inactive' state of the immunological homunculus, and onto the cognitive HPA axis, determining both onset and progression of autoimmune disease. The (relatively) protective role of female hormones against CHD, given the indistinguishability of men and women in figure 1, implies existence of a plastic, pleiotropic, response of the immune system and HPA axis to PSH. In essence, one has a sex-based choice of death by hanging or by firing squad, i.e. CHD induced by chronic vascular inflammation for African-American men, or a particular induced autoimmune disease for African-American women, SLE. A roughly similar story can probably be told regarding the increased rate of aggressively fatal breast cancer, diabetes, and other disorders in African-American women.

That is, the 'message' of PSH in the US is written onto the bodies of African-American men and women as, respectively, elevated rates of coronary heart disease, systemic lupus erythematosus, and allied disorders of chronic inflammation.

The rise in SLE among African-American women appears to parallel the rise of asthma among US minority urban children, which has increased 50 percent since 1980 (CDC,, 1996; NCHS, 1996). As we and others have described, (Wallace, Wallace and Fullilove, 2002; Carr et al., 1992) the geography of asthma in places like New York City closely matches the geography of public policy-driven urban burnout, contagious urban decay, and 'urban renewal' which has left most US urban minority neighborhoods looking like Dresden after the firebombing. Elsewhere (Wallace, Wallace and Fullilove, 2002) we have interpreted the rise of asthma among urban minority children as the writing of a kind of deliberate community lynching upon the developing immune system. Geographic analysis might well show that rising rates of SLE among African-American women represent the writing of that practice upon the developed immune system of women of reproductive age. In both cases a Th2 phenotype appears to be imposed. These are questions for future research.

As many have argued, health disparities are inevitably only the tip of an iceberg which can enmesh powerful or majority populations into dynamics affecting the marginalized.

Relative raised rates of autoimmune disease among African-American women are a red flag: Pathogenic social hierarchy may place a severe biological limit on the ultimate effectiveness of traditional medical behavioral and drug approaches to immune-related disease across all US subgroups, not merely for African-Americans.

This suggests in particular that 'magic bullet' medical interventions against lupus, to be effective at the population level, must be integrated as part of a larger 'ecosystem' strategy addressing the more basic problems of pathogenic social hierarchy and gender discrimination in the US. African-American women are, however, doubly burdened through the synergism of historical patterns of racism with a traditional gender discrimination which may, in fact, reflect that racism within African-American communities, and should be among the first to benefit from such reforms.

The remarkable rise of both lupus and asthma in US minority communities after 1980 seems to indicate, from this perspective, the tightening of discrimination rather than any efforts at reform. Our own studies (e.g. D. Wallace and R. Wallace, 1998; R. Wallace and D. Wallace, 1997) suggest that the inevitable failure of American Apartheid to effectively shield the powerful from the forces and impacts of marginalization means that the dominant population is being brought into a dynamic of increasing pathology as well. Nobody is more entrained into systems like figure 1 than the white majority in the US, which holds itself within the same structure it holds others, and would thus benefit by reform. Such, indeed, was the message of the Rev. Dr. Martin Luther King Jr., a message which appears to have a very basic biological reality.

## Acknowledgments

This research was partially supported through NIEHS Grant I-P50-ES09600-04, and benefited from earlier monies provided under an Investigator Award in Health Policy Research from the Robert Wood Johnson Foundation. The author thanks Dr. Noel Rose and an anonymous reviewer for remarks useful in revision.

## Figure caption

**Figure 1.** Redisplay of data from Singh-Manoux et al. (2003). Sex-specific dose-response curves of age-adjusted prevalence of self-reported ill-health vs. self-reported status rank, Whitehall II cohort, 1997 and 1999. 1 is high and 10 is low status. Note that the curves are virtually identical, and that the upper point is very near the EC-50 level in this population. Self-reported health is a highly significant predictor of later morbidity and mortality.

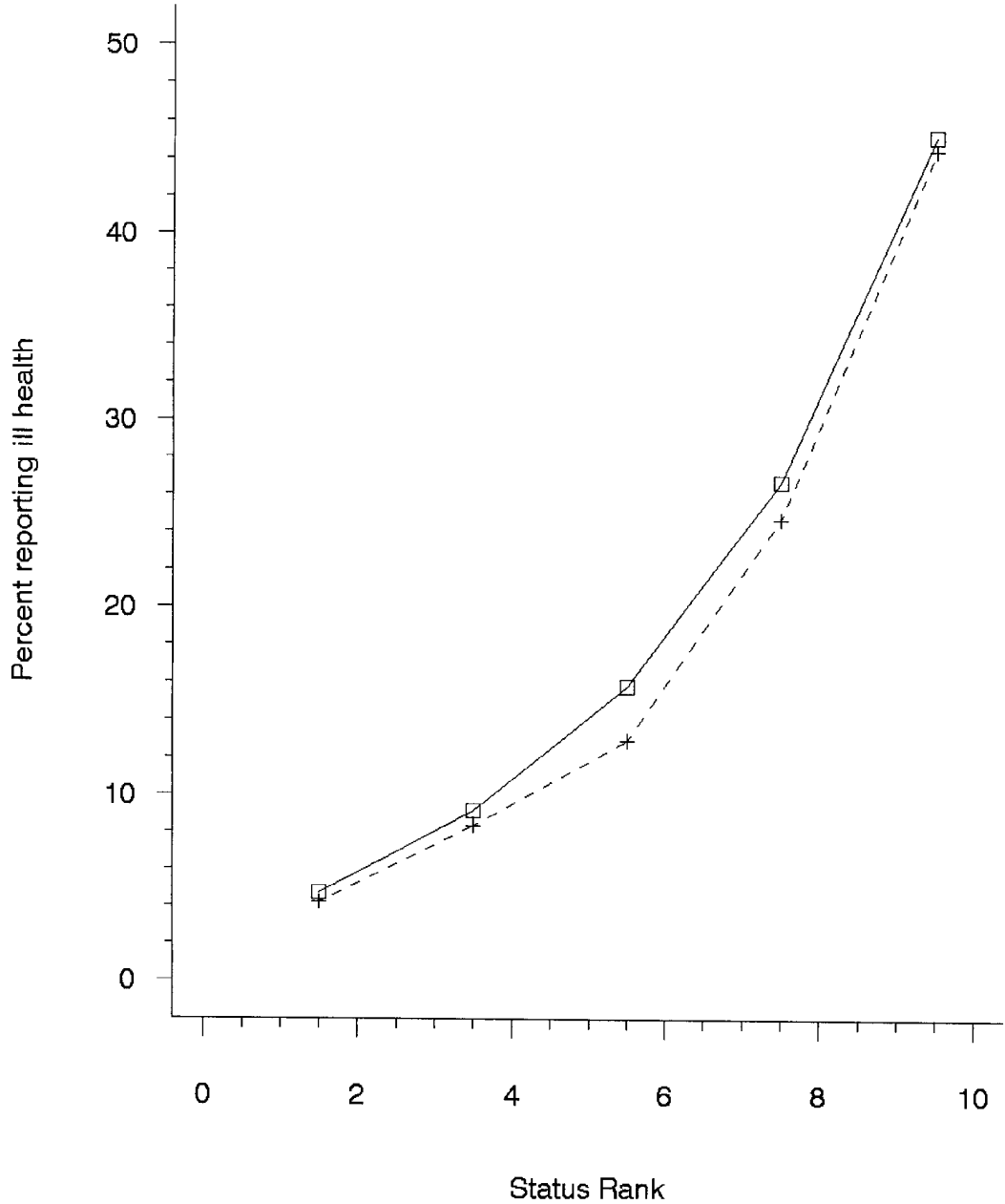